\documentclass[twocolumn,showpacs,amsmath,amssymb,superscriptaddress]{revtex4}

\usepackage{dcolumn}

\usepackage{graphicx}
\usepackage{dcolumn}
\usepackage{bm}

\begin{document}


\title{Structural evolution in Pt isotopes with the Interacting 
Boson Model Hamiltonian derived from the 
Gogny Energy Density Functional }

\author{K.~Nomura}
\affiliation{Department of physics, University of Tokyo, Hongo,
Bunkyo-ku, Tokyo, 113-0033, Japan} 

\author{T.~Otsuka}
\affiliation{Department of physics, University of Tokyo, Hongo,
Bunkyo-ku, Tokyo, 113-0033, Japan} 
\affiliation{Center for Nuclear Study, University of Tokyo, Hongo,
Bunkyo-ku Tokyo, 113-0033, Japan} 
\affiliation{National Superconducting Cyclotron Laboratory, 
Michigan State University, East Lansing, MI}

\author{R.~Rodr\'\i guez-Guzm\'an}
\affiliation{Instituto de Estructura de la Materia, CSIC, Serrano
123, E-28006 Madrid, Spain} 

\author{L.~M.~Robledo}
\affiliation{Departamento de F\'\i sica Te\'orica, Universidad
Aut\'onoma de Madrid, E-28049 Madrid, Spain}

\author{P.~Sarriguren}
\affiliation{Instituto de Estructura de la Materia, CSIC, Serrano
123, E-28006 Madrid, Spain}

\date{\today}

\begin{abstract}

Spectroscopic calculations are carried out, for the 
 description of the shape/phase transition in Pt nuclei in terms of the
 Interacting Boson Model (IBM) Hamiltonian derived 
 from (constrained) Hartree-Fock-Bogoliubov (HFB) calculations with the
 finite range and density dependent Gogny-D1S Energy Density Functional. 
Assuming that the many-nucleon driven dynamics of nuclear surface
 deformation can be simulated by effective bosonic degrees of freedom,
 the Gogny-D1S potential energy 
 surface (PES) with quadrupole degrees of freedom is mapped onto the
 corresponding PES of the IBM. 
Using this mapping procedure, the parameters of the IBM
 Hamiltonian, relevant to the low-lying quadrupole collective states, are
 derived as functions of the number of valence nucleons. 
Merits of both Gogny-HFB and IBM approaches are utilized so that the spectra
 and the wave functions in the laboratory system are calculated precisely. 
The experimental low-lying spectra of both ground-state and side-band
 levels are well reproduced. 
From the systematics of the calculated spectra and the reduced E2
 transition probabilities $B$(E2), the prolate-to-oblate shape/phase
 transition is shown to take 
 place quite smoothly as a function of neutron number $N$ in the
 considered Pt isotopic chain, for which the $\gamma$-softness plays an
 essential role. 
All these spectroscopic observables behave consistently with the
 relevant PESs and the derived parameters of the IBM Hamiltonian as
 functions of $N$. 
Spectroscopic predictions are also made for those nuclei which do not 
 have enough experimental E2 data. 

\end{abstract}


\maketitle



\section{Introduction\label{sec:introduction}}

The quadrupole collective motion has always attracted considerable
attention in nuclear physics \cite{BM,Collective,RS}. Nevertheless, a 
fully microscopic understanding of the evolution of the nuclear shapes 
with the number of nucleons still remains a major challenge
\cite{review,Bender_review,Heenen-nature,Werner,Robledo-1,Robledo-2,Rayner-1,rayner-PRL,Naza-def,Mottelson}. 
From the experimental point 
of view, low-lying spectroscopy is one of the 
most powerful sources of information about structural evolution and/or
shape transitions in atomic nuclei since it allows to establish
signatures correlating the excitation energies with deformation
properties \cite{Julin,draco,davidson,wu96,podolyak,pfun,caamano}. 
In particular, the complex interplay between several deformation degrees
of freedom, taking place in different regions of the nuclear chart,
offers the possibility of testing microscopic descriptions of atomic
nuclei under a wide variety of conditions. 
In this context, mean-field approximations based on 
effective Energy Density Functionals (EDFs), which are a cornerstone to
almost all microscopic approximations to the nuclear many-body problem
\cite{RS}, appear to be a first tool to rely on when looking for
fingerprints of nuclear shape/phase transitions.

Mean-field approximations are based on product trial 
wave functions, which are used to minimize a given EDF. 
Such products break several symmetries of the underlying nuclear
Hamiltonian (spontaneous symmetry breaking mechanism) allowing the use 
of an enlarged Hilbert space within which static correlations
associated with collective modes (e.g., quadrupole deformations) are
incorporated at the cost of a moderate effort. Nowadays, systematic
mean-field studies are possible because, on the one hand, important
advances have been made in the fitting protocols providing EDFs with
global predictive power all over the nuclear chart. 
Popular EDFs for calculations along these lines are the non-relativistic
Gogny \cite{Go,gogny-other} and Skyrme \cite{Bender_review,Sk,VB} ones,
as well as different parameterizations of the relativistic mean-field
Lagrangian \cite{Bender_review,Vetrenar-1}. On the other hand, it has
also become possible to recast mean-field equations in terms of
efficient minimization procedures such as the so-called gradient method
\cite{gradient-1,gradient-2}. 
One of the advantages of the gradient method 
is the way it handles constraints, which is well adapted to the case 
where a large number of constraints are required (like the case 
which requires, in addition to the proton and neutron number 
constraints, constrains on both $\beta$ and $\gamma$
degrees of freedom characterizing the nuclear shape). 
Another advantage is its robustness in reaching a solution, a convenient 
property when large scale calculations requiring the solution of many
HFB equations are performed.

On its own, the Interacting Boson Model (IBM) \cite{AI} has been quite 
successful in reproducing the experimental spectra and electromagnetic
transitions for low-lying quadrupole collective states. 
The virtue of the IBM is, with its simplicity, its robust
capability of calculating the spectroscopic observables precisely, while
the parameters of the IBM Hamiltonian have been determined
phenomenologically. 
Therefore, the IBM itself has a certain microscopic 
foundation, where the collective $J=0^+$ ($S$) and $2^+$ ($D$) pairs of
valence nucleons are approximated by $J=0^+$ ($s$) and $2^+$ ($d$)
bosons, respectively \cite{OAI}. 
The proton and neutron degrees of freedom can be taken into
account, where the so-called proton (neutron) $s_{\pi}$ and $d_{\pi}$
($s_{\nu}$ and $d_{\nu}$) bosons correspond to the collective pairs of 
valence protons (neutrons) $S_{\pi}$ and $D_{\pi}$ ($S_{\nu}$ and
$D_{\nu}$) \cite{OAI,OCas}. 
This is closer to a microscopic picture compared to a simpler version of 
IBM and is known as the proton-neutron Interacting Boson Model (IBM-2). 
As the number of valence protons (neutrons) is 
constant for a given nucleus, the number of proton (neutron) bosons, 
denoted by $n_{\pi}$ ($n_{\nu}$), is set equal to half of the 
valence proton (neutron) numbers. 
The derivation of the IBM Hamiltonian has been studied extensively in
realistic cases for nearly spherical or 
$\gamma$-unstable shapes \cite{MO,Deleze,seniority,GRS82} using 
generalized seniority states of the shell model \cite{OAI,OCas}, as well
as for deformed nuclei \cite{OPLB138,OY}, but still remains to be
done for general cases in a unified manner. 
Therefore, it is timely and necessary to bridge the IBM and mean-field 
models with the help of fermion-to-boson mapping procedures. 
The key question here is to investigate to which 
extent the underlying fermionic dynamics of mean-field models can be
translated into effective bosonic degrees of freedom. 
Such an approach would enable one to take advantage of the universality
of microscopic nuclear EDFs 
\cite{Go,gogny-other,Bender_review,Sk,VB,Vetrenar-1} and  
the simplicity of the IBM \cite{AI}. 
By the combination of both models, one would be able to access the
spectroscopic observables which have the good quantum numbers in the
laboratory system, including those for experimentally unexplored nuclei. 

A novel way of deriving the parameters 
of the IBM Hamiltonian has been recently proposed by two of us
\cite{nso}. 
The IBM Hamiltonian has been constructed by mapping the 
mean-field potential energy surface (PES), obtained in the framework of 
(constrained) Skyrme Hartree-Fock plus BCS calculations
\cite{ev8-code}, onto the corresponding PES of the IBM. 
The parameters of the IBM Hamiltonian, relevant to the description of
the considered quadrupole collective states, have been shown to be
determined uniquely as function of the number of valence nucleons by
using the Wavelet analysis \cite{nsofull}. 
Calculations along these lines have been performed 
to study the shape/phase transition in Sm isotopes with neutron number
$N=82\sim 96$, as well as for Ba, Xe, Ru and Pd isotopes with
$N=50\sim 82$. Spectroscopic predictions have also been made for 
W and Os nuclei with $N>$126 in the lower-right quadrant of $^{208}$Pb
\cite{nso,nsofull}. 
In addition, it has to be mentioned that the quantum mechanical
correlation effect on the binding energies can be included in such
calculations by diagonalizing the mapped IBM Hamiltonian
\cite{nsofull}. 
Note that, in this framework, the IBM keeps its important properties
including the boson number counting rule as well as the algebraic
features. 

In this paper we present, spectroscopic calculations 
for the Pt isotopic chain (i.e., for the even-even isotopes $^{172-200}$Pt)
in terms of an IBM Hamiltonian determined 
microscopically by mapping the  PES obtained in the framework of  
the (constrained) Hartree-Fock-Bogoliubov (HFB) approximation 
\cite{Robledo-2,rayner-PRL,gradient-2}
based on the 
parametrization D1S \cite{D1S} of the Gogny-EDF \cite{Go,gogny-other}.
Quite recently, the structural evolution in Pt isotopes, including the 
role of triaxility (i.e., the $\gamma$ degree of freedom), has been studied by 
three of us \cite{RaynerPt}. 
In addition to the (standard) Gogny-D1S EDF, the new incarnations 
D1N \cite{D1N} and D1M \cite{D1M} of the Gogny-EDF have also been
included in the mean-field analysis of Ref. \cite{RaynerPt}.  
The considered range of neutron numbers included prolate, triaxial,
oblate and spherical shapes and served for a detailed comparison of the
(mean-field) predictions of the new parameter sets D1N and D1M against
the standard parametrization D1S. 
It has been shown that, regardless of 
the particular version of the Gogny-EDF employed, the prolate-to-oblate 
shape/phase transition occurs
quite smoothly  with the $\gamma$-softness playing an  important
role. 
It is therefore very interesting to study how the systematics of
the HFB PESs discussed in Ref. \cite{RaynerPt} is reflected in the
isotopic evolution of the corresponding low-lying quadrupole collective
states and how accurately such states can be reproduced by a mapped IBM
Hamiltonian \cite{nso,nsofull}. 
Let us stress that our main goal in the
present work is to study the performance of a fermion-to-boson mapping
procedure \cite{nso,nsofull} based on the Gogny-EDF. For this reason, as
a first step, we will restrict ourselves to a mapping in terms of the
parametrization Gogny-D1S already considered as global and able to
describe reasonably well low-energy experimental data all over the
nuclear chart (see, for example, Refs.  \cite{gradient-2,CollGo} and  
references therein).

From the theoretical perspective, the Pt and  neighboring 
isotopic chains have been extensively studied in terms of both IBM 
and mean-field-based approaches. 
There is much experimental evidence \cite{Pt196O6,CastenCizewski}
revealing existences of $\gamma$-unstable O(6) nuclei in Pt 
isotopes. 
The IBM-2 has been used in a phenomenological way for the spectroscopy
of Pt, Os and W isotopes \cite{DB_W,BijkerOs}. 
The prolate-to-oblate transition in Pt as well as in Os and W 
nuclei, has been observed in the recent experiment \cite{198Os}, where a
relatively moderate oblate-to-prolate shape/phase 
transition occurs in Pt as compared to Os and W nuclei. 
Spectroscopic calculations have been carried out for Pt isotopes in the 
framework of the five-dimensional collective Hamiltonian, derived from
the pairing-plus-quadrupole model \cite{KummerBarangerPt}. 
Evidence for $\gamma$ vibrations and shape evolution in 
$^{184-190}$Hg has been considered in Ref.~\cite{Hg-Dela}, where a
five-dimensional collective Hamiltonian was built with the help of
constrained Gogny-D1S HFB calculations. 
On the other hand, systematic mean-field studies
of the evolution of the ground state shapes in Pt 
and the neighboring Yb, Hf, W and Os nuclei
have been carried 
out with non-relativistic Skyrme \cite{Robledo-2} and 
Gogny \cite{gradient-2,RaynerPt} EDFs, as well as within the 
framework of the relativistic mean-field (RMF) approximation
\cite{RMFPt}. 
One should also keep in mind, that Pt, Pb and Hg nuclei belong to a
region of the nuclear chart, around the proton shell closure $Z=82$, 
characterized by a  pronounced  competition between  low-lying
configurations corresponding to different intrinsic deformations
\cite{Andre} and therefore, a detailed description of the very rich
structural evolution in these nuclei requires the inclusion  
of correlations beyond the static mean-field picture 
\cite{DuguetGCMPb,BenderGCMPb,RaynerGCMPb}
accounting for both symmetry restoration and configuration mixing. 
The role of configuration mixing in this region has also been considered
in phenomenological IBM studies \cite{CMIBM,IBMCMECQF}. 

The paper is organized as follows. In Sec.~\ref{sec:theory}, we will
briefly describe the theoretical tools used in the present
study. 
Illustrative examples of IBM PESs, obtained by mapping the corresponding
Gogny-HFB PESs, are presented in Sec.~\ref{sec:IBM-PES}. 
The isotopic
evolution of the IBM parameters derived for the nuclei $^{172-200}$Pt
is discussed in Sec.~\ref{sec:IBM-PARAMETERS}. 
Spectroscopic calculations,
including the systematics of  excitation spectra and  reduced E2
transition probabilities $B$(E2) along the Pt isotopic chain, will
be discussed in Sec.~\ref{sec:results}. 
There, we will also show detailed comparisons
between the predicted level schemes and the available data for 
some Pt isotopes selected as a representative sample.
Finally, Sec.~\ref{sec:summary} is devoted to the concluding 
remarks and work perspectives.

\section{Theoretical procedure}
\label{sec:theory}

In this section, we briefly describe the theoretical frameworks 
used in the present study, i.e., the constrained HFB  approximation, as well as 
the procedure followed to construct
the corresponding mapped IBM Hamiltonian. For more details the reader is referred to 
Refs. \cite{gradient-2,RaynerPt} and \cite{nsofull}.

In order to compute the Gogny-HFB PESs, which are our starting point, we
have used the (constrained) HFB method together with the parametrization
D1S  of the Gogny-EDF. 
The solution of the HFB equations, leading to the set of vacua $|
\Phi_{\rm HFB} (\beta,\gamma)\rangle$, is based on the equivalence
of the HFB with a minimization problem that is solved using the 
gradient method \cite{gradient-1,gradient-2}. In agreement with the fitting
protocol of the force, the
kinetic energy of the center of mass motion has been subtracted from the
Routhian to be minimized in order to ensure 
that the center of mass is kept at rest. The exchange 
Coulomb energy is considered in the Slater approximation and
we neglect the contribution
of the Coulomb interaction to the pairing field. The HFB  quasiparticle
operators are expanded  
in a Harmonic Oscillator (HO) basis containing enough 
number of shells (i.e., $N_{shell}=13$ major shells) to grant 
convergence for all values of the mass quadrupole operators 
and for all the nuclei
studied. 
We constrain the average values of 
the mass quadrupole operators 
$\hat{Q}_{20}=\frac{1}{2}\left(  2z^{2} - 
x^{2} - y^{2}\right)$ and 
$\hat{Q}_{22}= \frac{\sqrt{3}}{2}\left(x^{2} - y^{2} \right)$ to the desired 
deformation values $Q_{20}$ and $Q_{22}$ defined as 

\begin{equation}
\label{Q20}
Q_{20}=\langle \Phi_{\rm HFB} | \hat{Q}_{20} | \Phi_{\rm HFB}
\rangle
\end{equation}
and 

\begin{equation}
\label{eq:Q22}
Q_{22}=\langle \Phi_{\rm HFB} | \hat{Q}_{22} | \Phi_{\rm HFB} 
\rangle. 
\end{equation}

In Ref.~\cite{RaynerPt},  the  $Q-\gamma$ 
energy contour plots with

\begin{equation}
\label{Q0}
Q = \sqrt{  Q_{20}^{2}+ Q_{22}^{2}}
\end{equation}
and 

\begin{equation}
\label{gamma-fermion}
\tan \gamma = \frac{Q_{22}}{Q_{20}}
\end{equation}
have been used to study the (mean-field) evolution of the 
ground state shapes in Pt nuclei. Alternatively, one could 
also consider the  $\beta-\gamma$ representation in which the
quadrupole deformation parameter 
$\beta$ is written \cite{gradient-2} in terms 
of $Q$ [Eq.(\ref{Q0})] as 
\begin{eqnarray}
\label{beta-fermions}
\beta = \sqrt{\frac{4\pi}{5}} \frac{Q}{A \langle r^{2} \rangle}
\end{eqnarray}
where $\langle r^{2} \rangle$ represents the mean squared radius evaluated
with the corresponding HFB state $ | \Phi_{\rm HFB} \rangle$. 

The set of constrained
HFB calculations described above, provides  the Gogny-D1S $\beta-\gamma$ PES 
(i.e., the total HFB energies $E_{\rm HFB}(\beta,\gamma)$ \cite{RS})
 required for 
the subsequent mapping procedure, 
for which the following IBM-2 Hamiltonian 
$\hat H_{\rm IBM}$
is employed
 
\begin{eqnarray}
\hat H_{\rm IBM} = \epsilon(\hat n_{d \pi}+\hat n_{d \nu})+\kappa
 \hat Q_{\pi}\cdot \hat Q_{\nu}. 
\label{eq:bh}
\end{eqnarray}
where 

\begin{eqnarray}
\hat n_{d \rho}=d_{\rho}^{\dagger}\cdot\tilde d_{\rho} , \quad (\rho =\pi,\nu)
\end{eqnarray}
and 

\begin{eqnarray}
\hat Q_{\rho}=[s_{\rho}^{\dagger}\tilde d_{\rho}+d_{\rho}^{\dagger}\tilde
s_{\rho}]^{(2)}+\chi_{\rho}[d_{\rho}^{\dagger}\tilde d_{\rho}]^{(2)}
\end{eqnarray}
stand for the $d$-boson number operator
and the quadrupole operator, respectively. The competition between the
coupling constants 
$\epsilon$ and $\kappa$ determines the degree of nuclear deformation.

The bosonic PES is represented by the 
expectation value of $\hat H_{\rm IBM}$, computed in terms of the
so-called boson coherent state \cite{DS,GK,BMc}

\begin{eqnarray} 
|\Phi\rangle\propto\prod_{\rho=\pi,\nu}\Big[s_{\rho}^{\dagger}+\sum_{\mu=0,\pm 2}\alpha_{\rho\mu}d_{\rho\mu}^{\dagger}\Big]^{n_{\rho}}|0\rangle   
\end{eqnarray}
where $|0\rangle$ stands for the boson vacuum (i.e., inert core) and the
coefficients $\alpha$'s are expressed as 
$\alpha_{\rho 0}=\beta_{\rho}\cos{\gamma_{\rho}}$, 
$\alpha_{\rho\pm 1}=0$ and 
$\alpha_{\rho\pm 2}=\frac{1}{\sqrt{2}}\beta_{\rho}\sin{\gamma_{\rho}}$. 
Within this context, the intrinsic shape of the nucleus is described in terms of the
(axially symmetric) deformation $\beta_{\rho}$ and the (triaxial) deformation
$\gamma_{\rho}$. 
In the present study, as well as in our previous works
\cite{nso,nsofull}, we assume for simplicity that
$\beta_{\pi}=\beta_{\nu}\equiv\beta_{\rm B}$ and
$\gamma_{\pi}=\gamma_{\nu}\equiv\gamma_{\rm B}$. 
The IBM PES is then given by \cite{nso,nsofull}

\begin{eqnarray}
& & E_{\rm IBM}(\beta_{\rm B},\gamma_{\rm B})=
\frac{\epsilon(n_{\pi}+n_{\nu})
 \beta_{\rm B}^2}{1+\beta_{\rm B}^2}+n_{\pi}n_{\nu}\kappa
 \frac{\beta_{\rm B}^2}{(1+\beta_{\rm B}^2)^2}\times \nonumber \\
& &\Big[4-2\sqrt{\frac{2}{7}}(\chi_{\pi}+\chi_{\nu})\beta_{\rm 
  B}\cos{3\gamma_{\rm B}}+\frac{2}{7}\chi_{\pi}\chi_{\nu}\beta_{\rm
  B}^2\Big]. 
\label{eq:IBM-PES}
\end{eqnarray}

Here we assume the proportionality 
$\beta_{\rm B}=C_{\beta}\beta$, with $C_{\beta}$ being a numerical
coefficient \cite{nso}. 
If one further assumes the separability of the mapping  
along the $\beta$ and $\gamma$ directions \cite{nso,nsofull}, one then
has $\gamma_{\rm B}=\gamma$. 
Thus, ($\beta_{\rm B}$, $\gamma_{\rm B}$) represent the boson images of  
the (fermion) deformation parameters ($\beta$, $\gamma$) given by
Eqs.~(\ref{gamma-fermion}) and (\ref{beta-fermions}). 
We then map a point on the HFB PES, ($\beta$,$\gamma$), within an energy
range relevant for the considered low-lying quadrupole collective
states, onto the corresponding point on the IBM PES, 
($\beta_{\rm B}$,$\gamma_{\rm B}$). 
This process is exactly the mapping of the fermionic PES onto the
bosonic one. 
In practice, one determines the $\epsilon$, 
$\kappa$, $\chi_{\pi,\nu}$ and $C_{\beta}$ values for each individual
nucleus by drawing the IBM PES so that the topology of the corresponding
HFB PES is reproduced. 
This is done unambiguously by means of the recently developed procedure
\cite{nsofull}, which makes use of the powerful method of the Wavelet transform
\cite{wavelet}. 

Here we would like to make the following remarks: 
The topology of the HFB PES reflects essential (fermionic) features of 
many-nucleon systems, such as the Pauli principle and the 
underlying nuclear interactions. Such effects are supposed to be
incorporated into the boson system by the mapping procedure
\cite{nso}. On the other hand, the solution of the five-dimensional (5D)
collective Bohr Hamiltonian with  
parameters obtained from  EDFs calculations (see, for 
example, \cite{CollSk,CollGo,CollRHB}) is a popular alternative to
obtain the low-lying collective spectra in even-even nuclei. In this
kind of calculations the pure mean-field PES is replaced by another
quantity that incorporates in addition to the HFB energy, the zero point
rotational and vibrational corrections. These corrections to the energy
are intimately related to the use of a generalized kinetic energy term
for the collective motion that includes not only moments of inertia but
also collective vibrational masses. To what extent the present
mapping procedure plus the solution of the IBM Hamiltonian is able to
mimic the solution of the 5D Bohr Hamiltonian is still an open question,
that can be partially answered by looking at the reasonable results
obtained with our method and that compare qualitatively well with the 
ones of the 5D Hamiltonian. 
A possible way to incorporate the effect of the collective masses into
the mapping would be to make a change of variables (analogous to the
one invoked in the derivation of the GOA \cite{RS})
as to render the collective
masses constant all over the range of allowed values of the $\beta$ and
$\gamma$ deformation parameters. 
Perhaps, this is the missing element that could correct some observed
\cite{nso,nsofull} 
systematic deviations of the IBM rotational spectra with respect to the 
experimental ones for well deformed systems, and requires the introduction of
an additional mass term in the IBM Hamiltonian known as the  $L\cdot L$ term
\cite{ibmmass}. However, this problem does not show up for moderately deformed
cases like the ones studied in the present work and therefore 
it is not considered here. 

\section{Mapped potential energy surfaces}
\label{sec:IBM-PES}

\begin{figure}[ctb!]
\begin{center}
\includegraphics[width=8.5cm]{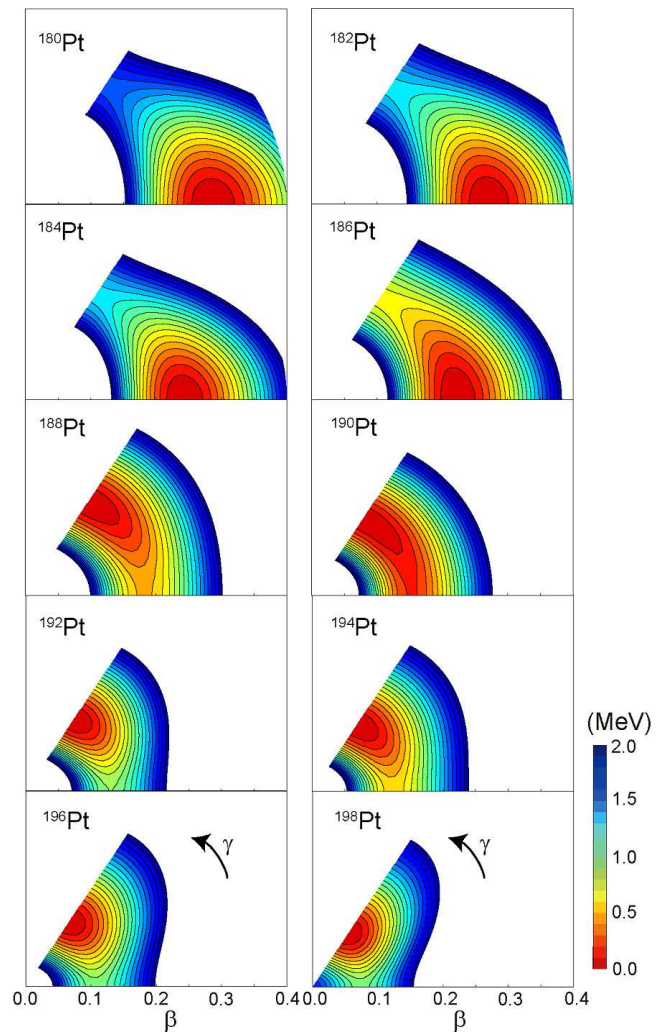}
\caption{(Color online) IBM PESs in $\beta\gamma$ plane for the nuclei
 $^{180-198}$Pt. Here, $\gamma=\gamma_{B}$ and $\beta= \beta_{B}/C_{\beta}$. 
The PESs are shown within $0.00\leqslant\beta\leqslant 0.40$ and
 $0^{\circ}\leqslant\gamma\leqslant 60^{\circ}$ up to 2 MeV excitation 
 from the minimum. Contour spacing is 100 keV. For details, see the 
 main text.}
\label{fig:pes}
\end{center}
\end{figure}

The IBM PESs obtained for the nuclei $^{180-198}$Pt are 
shown in Fig.~\ref{fig:pes}
as a representative sample. The IBM parameters
$\epsilon$,
$\kappa$, $\chi_{\pi,\nu}$ and $C_{\beta}$, to be discussed later on
in Sec.~\ref{sec:IBM-PARAMETERS}, have been obtained by mapping the
corresponding Gogny-D1S PESs presented in Fig.~2 of Ref.~\cite{RaynerPt}
along the lines previously described in Sec. \ref{sec:theory}. 

The IBM PESs from $^{180}$Pt to $^{186}$Pt
display a prolate deformed 
minimum and an oblate deformed saddle point. The prolate minimum becomes
softer in $\gamma$ but steeper in $\beta$ direction as the number of 
neutrons increases.
This is, roughly speaking, consistent with the topologies of the HFB
PESs of Ref.~\cite{RaynerPt}, where the minima are located a bit off
but quite nearby the line  $\gamma=0^{\circ}$. 

The IBM PESs for both $^{188,190}$Pt are 
$\gamma$ soft, having
the minimum on the oblate side. These nuclei are supposed to be 
close to the critical point of the prolate-to-oblate
shape  transition. The corresponding HFB PESs display shallow triaxial
minima with $\gamma\sim 30^{\circ}$ and are also soft along the 
$\gamma$ direction \cite{RaynerPt}. 
The IBM Hamiltonian considered in the present study does not provide a
triaxial minimum, but either 
prolate or oblate minimum, as can be seen from Eq.~(\ref{eq:IBM-PES}). 
The $\gamma$-softness can be simulated  by choosing the parameters 
$\chi_{\pi}$ and $\chi_{\nu}$ so that their sum becomes nearly equal to
zero. 
This is reasonable when a triaxial minimum is not deep enough like the
present case, where the triaxial minimum point in the HFB PES differs by
at most several hundred keV in energy from either prolate or oblate
saddle point.  
However, the topology of the mapped IBM PES is then somewhat sensitive to
the values of the parameters $\chi_{\pi}$ and $\chi_{\nu}$, which 
occasionally results in a quantitative difference in the location of
the minimum in the IBM PES from that of the HFB PES. 
In fact, and contrary to what happens with the HFB PESs \cite{RaynerPt},
the IBM PES of $^{190}$Pt is softer in $\gamma$ than that of
$^{188}$Pt. 
One should then expect a certain deviation of the resultant IBM spectra
from the experimental ones, which can be partly attributed to the small
difference already mentioned.

In Fig.~\ref{fig:pes}, isotopes from $^{192}$Pt to $^{198}$Pt exhibit
oblate deformation. 
The locations of their energy minima and their curvatures in both $\beta$
and $\gamma$ directions agree well with the ones of the Gogny-D1S PESs
in Ref. \cite{RaynerPt}. 
These isotopes become steeper in the 
$\gamma$ direction and shallower in the $\beta$ direction as the number
of neutrons increases. 
Their energy minima approach the origin more rapidly than the lighter Pt
nuclei shown Fig.~\ref{fig:pes}. 
This evolution reflects the transition from oblate deformed
ground states to a spherical vibrator
as one approaches the neutron shell closure $N=126$. 

\section{Derived IBM parameters}
\label{sec:IBM-PARAMETERS}

\begin{figure}[ctb!]
\begin{center}
\includegraphics[width=8.5cm]{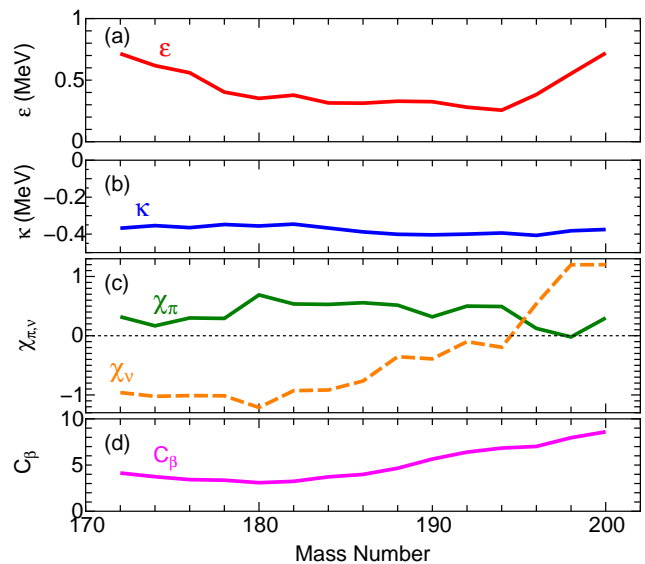}
\caption{(Color online) IBM parameters 
$\epsilon$, $\kappa$, $\chi_{\pi,\nu}$ and $C_{\beta}$ as
 functions of the mass  number A. For the wavelet analysis the
 Morlet function is used \cite{nsofull}. }
\label{fig:para}
\end{center}
\end{figure}

The IBM parameters
$\epsilon$,
$\kappa$, $\chi_{\pi,\nu}$ and $C_{\beta}$ derived for the nuclei
$^{172-200}$Pt from  the mapping procedure described 
in Sec. \ref{sec:theory} are depicted in
Figs.~\ref{fig:para}(a)-\ref{fig:para}(d) as functions of the mass 
number $A$.

Figure~\ref{fig:para}(a) shows the parameter $\epsilon$ gradually
decreases toward mid shell 
in accordance with the growth of the deformation. 
This trend reflects 
the structural evolution from nearly spherical to
more deformed shapes and is consistent with previous results
for other isotopic chains \cite{nsofull}. 
In Fig.~\ref{fig:para}(b), the derived $\kappa$ parameter is almost
constant and somewhat larger in comparison with the 
phenomenological value \cite{BijkerOs}, which is the consequence
of the sharp potential valleys observed in the Gogny-D1S PESs \cite{RaynerPt}.

On the other hand, in Fig.~\ref{fig:para}(c) the proton parameter
$\chi_{\pi}$ is almost constant  
while the neutron parameter $\chi_{\nu}$ changes significantly. 
The systematic behavior of the present $\chi_{\nu}$ value is consistent
with the phenomenological one \cite{BijkerOs}, while there is
quantitative difference between the former and the latter.  
The magnitude of the 
sum $\chi_{\pi}+\chi_{\nu}$ as well as its sign depend 
on how sharp the HFB PES is in the $\gamma$ direction 
and on whether the nucleus is prolate (negative sum) or oblate 
(positive sum) deformed, respectively. 
Therefore, as $\chi_{\pi}$ does not change much, the role of $\gamma$
instability can be seen clearly from the systematics of $\chi_{\nu}$. 
For the isotopes $^{172-180}$Pt the 
PES exhibits prolate deformation and the sum $\chi_{\pi}+\chi_{\nu}$ has 
negative sign. 
The average of the derived $\chi_{\pi}$ and $\chi_{\nu}$ values is
nearly equal to zero for the nuclei $^{182-194}$Pt. 
This is a consequence of the $\gamma$ softness in the corresponding HFB
PESs. 
On the other hand, the sum $\chi_{\pi}+\chi_{\nu}$ 
becomes larger with positive sign as we approach the neutron shell 
closure $N$=126 reflecting the appearance of weakly deformed oblate
structures in the corresponding PESs.

Figure~\ref{fig:para}(d) shows that $C_{\beta}$ decreases gradually
toward the middle of the major shell. 
$C_{\beta}$ can be interpreted as the ``bridge'' between
the geometrical deformation $\beta$ \cite{BM} and the IBM deformation
$\beta_{\rm B}$ and is thus proportional to the ratio between the total
and valence nucleon numbers, in a good approximation \cite{GK}. 
This is probably the reason 
for the decreasing trend observed in Fig.~\ref{fig:para}(d), as well as in
earlier studies for other isotopic
chains \cite{nso,nsofull}. 

\section{Spectroscopic calculations\label{sec:results}}

With all the parameters $\epsilon$,
$\kappa$, $\chi_{\pi,\nu}$ and $C_{\beta}$ required by the 
IBM Hamiltonian at hand, we are now able to test
the spectroscopic quality of our mapping procedure, based on the
Gogny-D1S EDF, for the nuclei $^{172-200}$Pt. Therefore, in the 
following we will discuss our predictions concerning the properties of the low-lying spectra
as well as the 
reduced transition probabilities $B$(E2). We will also consider 
their correspondence with
the mapped PESs and  the derived IBM parameters. We will compare our theoretical
predictions with the available 
experimental data taken from Brookhaven National Nuclear Data Center
(NNDC) \cite{data} and from the latest Nuclear Data Sheets \cite{BE2}.
The diagonalization of the IBM Hamiltonian is performed 
numerically for each nucleus using the code NPBOS \cite{npbos}. 

Here we have to note that the experimental $2^+_3$ and $4^+_3$ levels
for mass numbers $192\leqslant A\leqslant 200$ belong to bands different from that
of the $0^+_2$ level, while they are assigned to the quasi-$\beta$-band
levels in Fig.~\ref{fig:spectra}(b), as well as in
Fig.~\ref{fig:LevelScheme}, for convenience sake. 
Similarly, as one will see in Fig.~\ref{fig:spectra}(c) the experimental
data for the $3^+_1$ and the $4^+_2$ levels in $^{198,200}$Pt are
assigned to the quasi-$\gamma$-band levels lying on top of the $2^+_2$
energy. 

\subsection{Evolution of low-lying spectra}

\begin{figure*}[ctb!]
\begin{center}
\includegraphics[width=17cm]{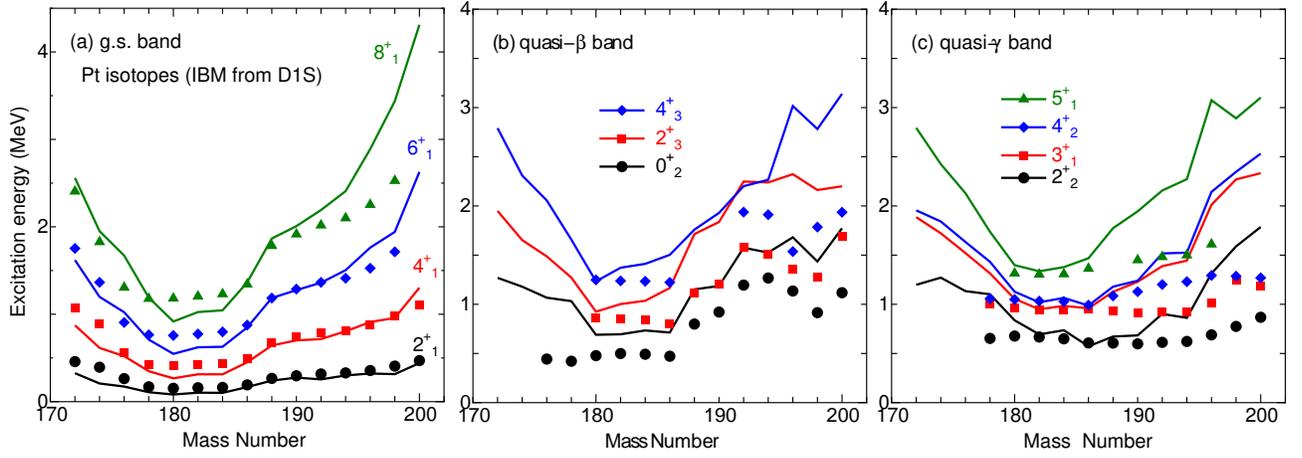}
\caption{(Color online) 
Evolution of calculated (curves) and experimental (symbols) low-lying
 spectra of $^{172-200}$Pt nuclei for (a) ground-state, (b) 
 quasi-$\beta$ and (c) quasi-$\gamma$ bands as functions of the mass
 number $A$. 
Experimental data are taken from Ref.~\cite{data}} 
\label{fig:spectra}
\end{center}
\end{figure*}

Figure~\ref{fig:spectra} displays
the calculated spectra for (a) ground-state, (b) quasi-$\beta$ and
(c) quasi-$\gamma$ bands. 
What is striking is the good agreement between the present calculations 
and the experimental data not only for ground-state but
also for quasi-$\beta$ and quasi-$\gamma$ band energies, where overall 
experimental trends are reproduced fairly well in particular for 
the open-shell nuclei $^{180-192}$Pt. 

We show in Fig.~\ref{fig:spectra}(a) the evolution of the 
$2_{1}^{+}$, $4_{1}^{+}$, $6_{1}^{+}$ and $8_{1}^{+}$
levels in the considered Pt nuclei as functions of the mass number $A$.  
The calculated energies decrease toward the middle of the major shell
with the number of the valence neutrons and remain almost constant 
for $176\lesssim A\lesssim 186$ nuclei. 
Although these tendencies are well reproduced, the rotational features
are somewhat enhanced in the calculated levels
for $^{180,182,184}$Pt which are slightly lower in energy than the 
experimental ones. From both the theoretical results and the
experimental data, one can also observe clear fingerprints for
structural evolution with a jump between $^{186}$Pt and $^{188}$Pt,
which can be correlated with the change of the mapped PESs from prolate
to oblate deformations. 
For $A\geqslant 188$ the yrast levels gradually go up as the neutron shell
closure $N$=126 is approached. 

One can also find signatures for a shape/phase transition in the 
systematics of the quasi-$\beta$ band levels shown in
Fig.~\ref{fig:spectra}(b). 
From $A=$180 to 186, the  $0^+_2$ band head and the $2^+_3$ level 
look either constant or nearly constant in both theory and experiment.  
The two levels are pushed up rather significantly from $A$=186 to 188 
consistently with the systematics in the ground-state band and with the
change of the mapped PESs as functions of the neutron number $N$.
The calculated $0^+_2$ and
$2^+_3$ levels are higher than but still follow the experimental trends.
 
Coming now to the quasi-$\gamma$ band levels shown in 
Fig.~\ref{fig:spectra}(c), one can observe the remarkable agreement 
between theoretical and experimental spectra for $180\leqslant A\leqslant 186$, 
where the $3^+_1$ level lies close to the $4_2^+$ level. 
However, the present calculation suggests this trend persists even 
for $188\leqslant A\leqslant 196$, whereas the 
relative spacing between the experimental $3^+_1$ and $4^+_2$ levels for 
these nuclei is larger. 
Similar deviation occurs for $5^{+}_{1}$ and $6^{+}_{2}$ levels, 
although the latter is not exhibited in Fig.~\ref{fig:spectra}(b). 
This means that our calculations suggest the feature characteristic of
the O(6) symmetry, where the staggering occurs as $2_{\gamma}^{+}$, 
($3_{\gamma}^{+}$ $4_{\gamma}^{+}$),
($5_{\gamma}^{+}$ $6_{\gamma}^{+}$), .... etc. 
However, the experimental levels are lying more 
regularly particularly for $188\leqslant A\leqslant 196$, and thus
appear to be in between the O(6) limit and a rigid triaxial rotor where
the staggering shows up as 
($2_{\gamma}^{+}$ $3_{\gamma}^{+}$), 
($4_{\gamma}^{+}$ $5_{\gamma}^{+}$), ... etc \cite{triaxial}. 
Such a deviation of the $\gamma$-band structure seems to be nothing 
but a consequence of an algebraic nature of the IBM, and indeed has also 
been found in existing phenomenological IBM calculations \cite{BijkerOs}. 
From a phenomenological point of view, the so-called cubic (or the
three-boson) interaction \cite{cubic,Casten-cubic} has been useful for
reproducing the experimental $\gamma$-band structure. 
The cubic term produces a shallow triaxial minimum that is seen in the 
Gogny-HFB PES, and may be introduced also in the Hamiltonian in
Eq.~(\ref{eq:bh}). 
This is, however, out of focus in the current theoretical framework,
because the cubic term represents an effective force whose origin
remains to be investigated further. 

Here, the deviations observed in the side-band levels (even in some of
the ground-state band levels) for $A\geqslant 196$ are probably related to
the larger magnitude of the parameter $\kappa$ as compared with its 
phenomenological value \cite{BijkerOs}. 
Roughly speaking, when the magnitude of $\kappa$ becomes larger, the
moment of inertia decreases, resulting in the deviation of not only
ground-state-band but also the side-band energies.  
The problem arises in the present case partly because, in the vicinity
of the shell closure $N=126$, 
the HFB PESs exhibit weak oblate deformations close to the origin
$\beta=0$, as we showed in Fig.~\ref{fig:pes}. 
In addition, the curvatures along the $\beta$ direction around the
minima are somewhat larger. 
These peculiar topologies of the Gogny-D1S PESs make it rather difficult 
to determine a value of $\kappa$ which gives reasonable agreement of
side-band energies with the experimental ones. 
In this case one may interpret that the deviation is mainly due to the 
properties of the particular version of the Gogny-EDF considered in the 
present study. 
Another possibility is that the boson Hamiltonian used may be still
simple, requiring the introduction of additional interaction terms in 
the boson system. 
Investigation along these lines is in progress and will be reported
elsewhere. 

\subsection{Systematics of $B$(E2) ratios}

\begin{figure}[ctb!]
\begin{center}
\includegraphics[width=8.5cm]{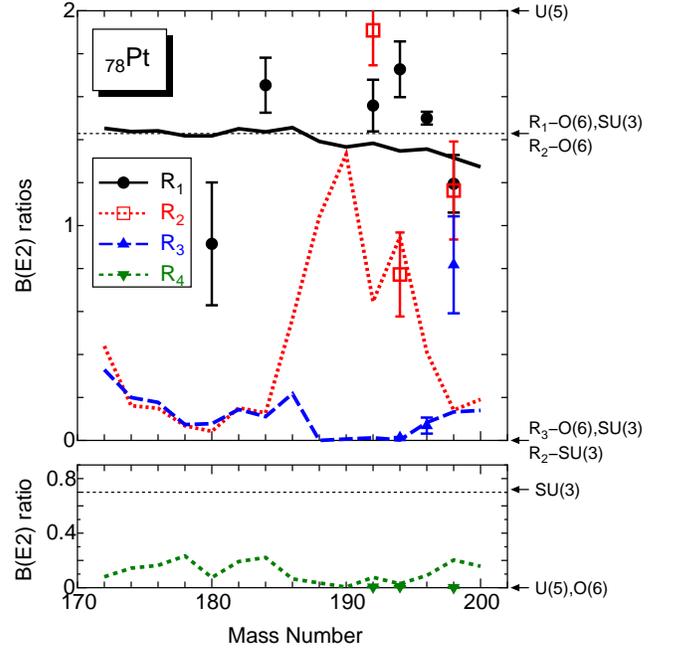}
\caption{(Color online) Reduced E2 transition probabilities $B$(E2) of
 $^{172-200}$Pt nuclei as functions of the mass number $A$.  
(a) $B$(E2) values for low-lying states normalized by
 $B$({\rm E2};$2^+_1\rightarrow 0^+_1)$ value, $R_1$, $R_2$ and $R_3$. 
(b) $B$(E2) branching ratio, 
$R_4=B$({\rm E2};$2^+_2\rightarrow 0^+_1)$/$B$({\rm E2};$2^+_2\rightarrow 2^+_1)$. 
Ratios $R_1-R_4$ are defined in Eq.~(\ref{eq:BE2}). 
Experimental data are taken from Ref. \cite{BE2}. For more details, see the 
main text.}
\label{fig:E2}
\end{center}
\end{figure}

Once the boson wave functions corresponding to the excited states 
of a given nucleus are obtained, we are able to compute electromagnetic
transitions, among which the reduced E2 transition 
probabilities $B$(E2) are of particular importance. The 
$B$(E2) transition probabilities are given by \cite{BM} 

\begin{eqnarray}
 B({\rm E2};J\rightarrow J^{\prime})=\frac{1}{2J+1}|\langle
  J^{\prime}||\hat T^{\rm (E2)}||J\rangle|^2, 
\end{eqnarray}
where $J$ and $J^{\prime}$ are the angular momenta for the initial and final
states, respectively. 
The E2 transition operator $\hat T^{{\rm (E2)}}$ is given 
by $\hat T^{\rm (E2)}=e_{\pi}\hat Q_{\pi}+e_{\nu}\hat Q_{\nu}$,
with $e_{\pi}$  and $e_{\nu}$ being the boson effective charges. 
In principle, the effective charges 
should be determined independently of the underlying  
mean-field calculation.  
In the present paper, we assume $e_{\pi}=e_{\nu}$, for simplicity, and 
focus our discussion on the $B$(E2) ratios defined as 

\begin{eqnarray}
R_1&=&B({\rm E2};4_1^+\rightarrow 2_1^+)/B({\rm
 E2};2_1^+\rightarrow 0_1^+) \nonumber \\
R_2&=&B({\rm E2};2_2^+\rightarrow 2_1^+)/B({\rm
 E2};2_1^+\rightarrow 0_1^+) \nonumber \\
R_3&=&B({\rm E2};0_2^+\rightarrow 2_1^+)/B({\rm
 E2};2_1^+\rightarrow 0_1^+) \nonumber \\
R_4&=&B({\rm E2};2_2^+\rightarrow 0_1^+)/B({\rm
 E2};2_2^+\rightarrow 2_1^+), 
\label{eq:BE2}
\end{eqnarray}
which are shown in Fig.~\ref{fig:E2} as functions of the mass number
$A$. 

The ratio $R_1$ is nearly constant all the way, being much below the
U(5) limit of IBM ($R_1=2$), and is rather close to $R_1=10/7$, which is 
the O(6) and SU(3) limit of IBM. 
Thus, $R_1$ is not a sensitive observable to distinguish between axially 
symmetric and $\gamma$-soft nuclei. 
This is reasonable because the structural
evolution between axially symmetric deformed and the $\gamma$-unstable
shapes is shown to take place quite smoothly from the systematics of the 
mapped PESs (in Sec.~\ref{sec:IBM-PES}) and the derived IBM parameters 
(in Sec.~\ref{sec:IBM-PARAMETERS}).   
The flat behavior of $R_1$ value for Pt isotopes in Fig.~\ref{fig:E2} 
differs from the one found e.g., in Sm isotopes \cite{IBM1-Sm}. 
There, a sharp decrease of $R_1$ value can be seen in the line of 
U(5)-SU(3) shape/phase transition. 

One can see that, in contrast to the flat systematics of the $R_1$ value
with respect to the mass number $A$, the ratio $R_2$ changes
significantly and is relatively large for $^{186-196}$Pt nuclei, being
close to $\frac{10}{7}$ (O(6) limit). 
This is consistent with the softness of the PESs for these
nuclei. 
Therefore, the quantity $R_2$ is quite sensitive to the shape 
evolution encountered in the PESs and can be thus considered as the best
signature for $\gamma$-softness among $R_1$-$R_4$.  
There are not much available data overall, but the experimental $R_2$
value is also relatively large around $^{192}$Pt. 
For the nuclei $^{176-184}$Pt, the theoretical $R_2$ value is close to
zero (the SU(3) limit) and slightly goes up from
$A$=174 to 172, probably approaching the U(5) vibrational limit 
($R_2$=2) in the vicinity of the neutron shell closure
$N=82$. 

Unlike the $R_2$ case, the calculated ratio $R_3$ does not change much
with mass number $A$ and is close to zero (O(6) and SU(3) limits of
$R_3$) for $^{188-196}$Pt. 
From $A$=180, the $R_3$ turns to increase 
as we move towards the neutron shell closure $N=82$ and is expected to
approach the U(5) limit ($R_3=2$). 
The calculated $R_3$ value is, however, still much smaller than the
experimental value at $A=198$. 
In fact, both the HFB and the mapped PESs for the nucleus 
$^{198}$Pt display a weakly deformed shape, which somewhat differs from
the vibrational feature expected from the corresponding experimental
levels. 
The present $R_3$ value does not exhibit a drastic change observed in
shape transitions in $A\sim$130 Ba-Xe and 
$A\sim$100 Ru-Pd isotopes, where the E2 transition from
the $0^+_2$ state to the $2^+_1$ is much enhanced \cite{nsofull}. 

Finally, the branching ratio $R_4$ also corresponds to a gradual shape
transition. The present calculations suggest that 
the $R_4$ value is nearly zero (O(6) limit) in the region where the
nuclei are soft and where the $R_2$ ratio takes large values. 
The calculated $R_4$ ratio becomes relatively larger for $A\leqslant 184$,
where the PESs show stronger prolate deformation.  
Consistently with the evolution of the IBM PESs, the calculated $R_4$
values turn to approach the U(5) limit, which is also zero, for
$A\leqslant 178$. 
Similarly to the $R_3$ case, a deviation from the vibrational character
of the experimental data is found at $A=198$. 
 
It should be emphasized that all the results for $B$(E2) values shown so
far are quite consistent with the topologies of the PESs and with the
derived IBM parameter values.

\subsection{Level schemes of selected nuclei}

\begin{figure*}[ctb!]
\begin{center}
\includegraphics[width=17cm]{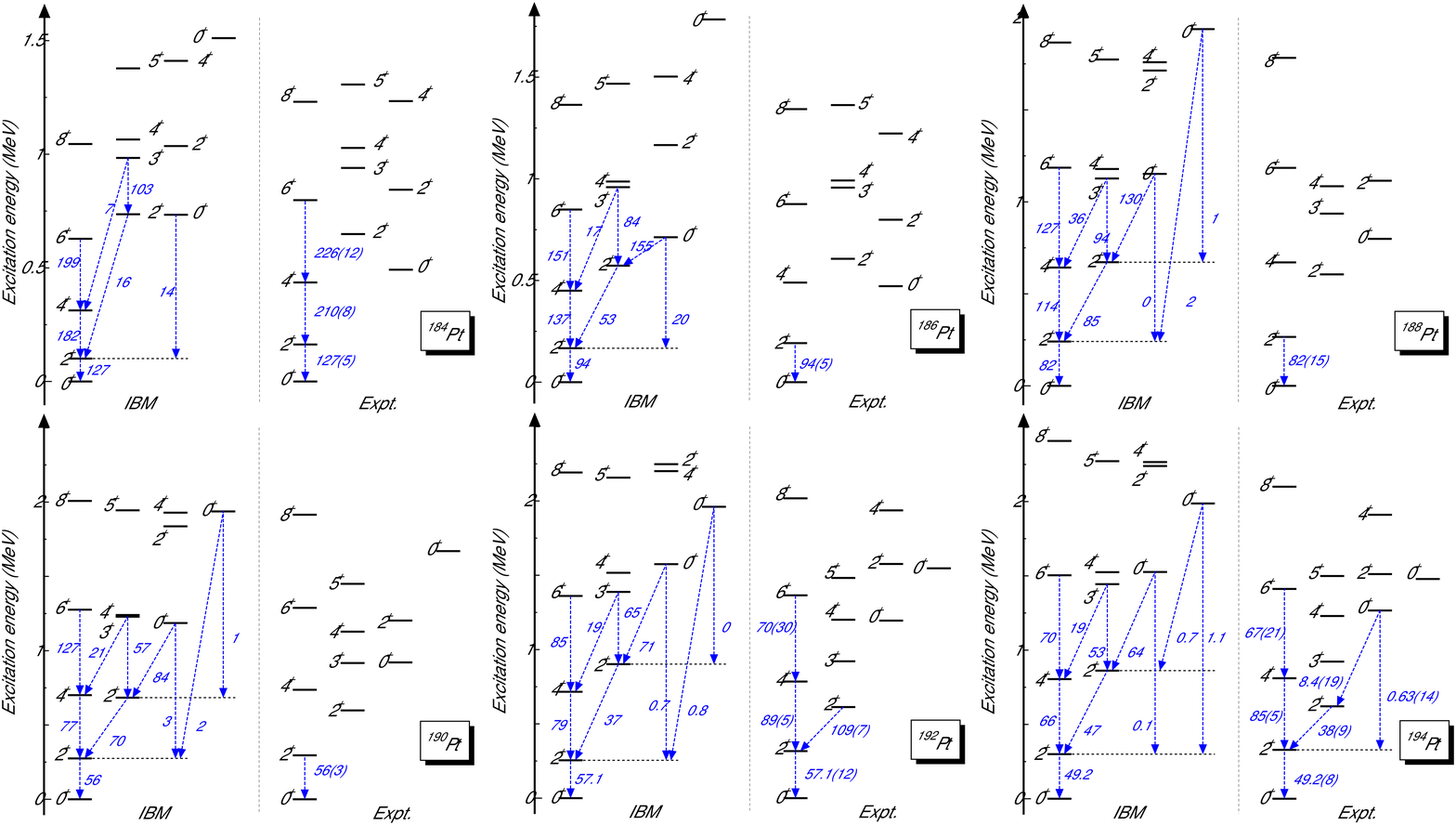}
\caption{(Color online) Level schemes for $^{184-194}$Pt nuclei. 
The Gogny D1S interaction is used. 
The theoretical $B$(E2) values (in Weisskopf units) for each nucleus are
 normalized to the experimental $B({\rm E2};2_1^+\rightarrow 0_1^+)$
 value. Note that, in $^{190}$Pt, the theoretical $3^+_1$ level is
 almost identical to but is lower by only 
 10 keV in energy than the $4^+_2$ level, where the E2
 transitions shown are those from the $3^+_1$ state, not from the
 $4^+_2$ state. For details, see the main text.}
\label{fig:LevelScheme}
\end{center}
\end{figure*}

As already mentioned in Sec.~\ref{sec:introduction}, one of the main
goals of the present study is to 
test the spectroscopic quality of the mapping procedure and the
underlying (universal) Gogny-D1S EDF \cite{gradient-2,CollGo}, which
have been described in Sec.~\ref{sec:theory}. 
Keeping this in mind, we will now turn our attention
to a more detailed comparison between our results and the 
available experimental data
for excitation spectra and $B$(E2) values. 
To this end, we select the nuclei $^{184-194}$Pt as a representative
sample corresponding to the mapped PESs shown in Fig.~\ref{fig:pes}. 
The level schemes obtained for the nuclei $^{184-194}$Pt are compared in  
Fig.~\ref{fig:LevelScheme} with available experimental data. 
The theoretical $B$(E2) values are shown also
in Fig.~\ref{fig:LevelScheme}. 
Note that $B({\rm E2};2_1^+\rightarrow 0_1^+)$ value is normalized to the 
experimental one. 
The virtue of the present calculation is to give predicted $B$(E2)
values for those nuclei which have no enough E2 information
available. 
This is particularly useful in the cases of $^{184}$Pt, $^{186}$Pt and
$^{188}$Pt where the calculated spectra agree well with the experiment.

For clarity, we divide the 
explanation of the results shown in Fig.~\ref{fig:LevelScheme} into 
the following three categories, according to the tendencies of what we
found in the PESs and the IBM parameters and the experimental data. 
The first is the prolate deformed regime represented by the nuclei
$^{184,186}$Pt, which exhibit a rotational character. 
Next, we will consider the isotopes $^{188,190}$Pt
which are apparently close to the critical point of the
prolate-to-oblate transition observed in the mapped PESs 
(see Fig.~\ref{fig:pes}). 
Lastly, calculated and experimental results are compared for the nuclei 
$^{192,194}$Pt which belong to the weakly  oblate deformed regime.  
Note that, in Fig.~\ref{fig:LevelScheme}, the energy scale is not common
for all nuclei. 

For $^{184,186}$Pt, the present calculation reproduces
overall pattern of the experimental spectra in all of the ground-state,
quasi-$\beta$ and quasi-$\gamma$ bands fairly well. 
Interesting enough, the 
bandhead energies, particularly the quasi-$\beta$ bandhead $0^{+}_{2}$,
are much higher than the $4^+_1$ level, compared to the experimental
data.  
This indicates that, reflecting the topologies of the PESs in
Fig.~\ref{fig:pes}, the $^{184,186}$Pt nuclei deviate from the
$\gamma$-soft O(6) character and exhibit rather rotational features. 
The $0^+_3$ energies are predicted to be above $4^{+}_{3}$ ones in both
nuclei. 

On the other hand, both $^{188}$Pt and $^{190}$Pt, whose PESs are quite
flat along the $\gamma$ direction in Fig.~\ref{fig:pes}, appear to be
closer to the $\gamma$-unstable O(6) limit of the IBM than the two
nuclei already mentioned above. 
The $2^+_2$ and $4^+_1$ levels lie close to each other in the
present study  and, in the spirit of group theory, are  
supposed to have the same $\tau=2$ quantum number of the
O(6) dynamical symmetry \cite{AI}. 
Similarly, in our calculations the $6^+_1$, $4^+_2$, $3^+_1$ and $0^+_2$ 
levels are almost degenerate 
and can be then grouped into the $\tau=3$ multiple. 
Along these lines, we can observe characteristic E2 decay patterns that
are quite consistent with the $\Delta\tau=\pm 1$ selection rule of the
O(6) limit \cite{AI}. 
For instance, the transition from the $0^+_2$ level (supposed to have
$\tau=3$) to $2^+_2$ level (supposed to have $\tau=2$) is dominant over
the one to $2^+_1$ (supposed to have $\tau=1$) in both $^{188}$Pt and
$^{190}$Pt.  
The trend characteristic of O(6) symmetry is clearly seen particularly
in $^{190}$Pt, where the sum of the parameters 
$\chi_{\pi}$ and $\chi_{\nu}$ almost vanishes as seen from
Fig.~\ref{fig:para}(c). 
This means that the nucleus is close to the pure O(6) limit, and is
consistent with the mapped PES in Fig.~\ref{fig:pes} that is nearly flat
along the $\gamma$ direction. 
Nevertheless, the structure of the corresponding experimental $\gamma$
band appears to have a more triaxial nature, where the $3^+_1$ 
and the $4^+_2$ levels are apart from each other. 
As we have anticipated in Sec.~\ref{sec:IBM-PES}, this deviation arises
partly due to the difference of the position of energy minimum between
Gogny-D1S PESs of \cite{RaynerPt} and the corresponding IBM PESs of
Fig.~\ref{fig:pes}. 

For $^{192,194}$Pt nuclei, the theoretical $\gamma$-band structure still
looks like that of O(6) symmetry. 
What is of particular interest here is that, for both $^{192,194}$Pt,
the relative location of the quasi-$\beta$-band head $0^{+}_{2}$ energy
is reproduced fairly well lying close to the $4^{+}_{2}$ level. 
In addition, for $^{194}$Pt, the present calculation suggests that the
$0^{+}_{2}\rightarrow 2^{+}_{2}$ E2 transition is dominant over the 
$0^{+}_{2}\rightarrow 2^{+}_{1}$ E2 transition, which, although there is
quantitative deviation, agrees with the experimental trend. 
The reason why such a quantitative difference occurs may be discussed
in the future. 
Compared to the experimental data, the theoretical 
quasi-$\gamma$ band is rather stretched and the band head $2^{+}_{2}$
energy is somewhat large. 
The calculated $0^{+}_{2}$ energy is also higher than the experimental
one in particular for $^{192}$Pt. 
Accordingly, the theoretical 
$B({\rm E2};2^{+}_{2}\rightarrow 2^{+}_{1})$ value is much 
smaller than experimental value with respect to the 
$B({\rm E2};2^{+}_{1}\rightarrow 0^{+}_{1})$ value. 
The deviations occur due to the derived $\kappa$ value, which is
somewhat larger than the phenomenological one \cite{BijkerOs}. 
For relatively high-lying side-band $2^{+}_{3}$ and
$4^{+}_{3}$ energies, the calculated results may not seem 
to be much reliable, because even the ordering of these levels are not
reproduced for $^{192}$Pt. 

\section{Summary\label{sec:summary}}

To summarize, spectroscopic calculations have been carried out, for the
Pt isotopic chain in terms of the Interacting Boson Model Hamiltonian
derived microscopically based on the (constrained)
Hartree-Fock-Bogoliubov approach with the Gogny-D1S Energy Density
Functional. 

The Gogny-HFB calculations provide the potential energy surface
(PES), which reflects, to a good extent, many-nucleon dynamics of
surface deformation with quadrupole degrees of freedom and structural
evolution in a given isotopic chain. 
By following the procedure proposed in Ref.~\cite{nso}, the PES of the
Gogny-D1S EDF is mapped onto the corresponding bosonic PES, and can be
then utilized as a guideline for  
determining the parameters of the IBM Hamiltonian. 
This enables one to calculate the spectroscopic observables with good
quantum numbers (i.e., the angular momentum and the particle number) in
the laboratory system without adjustment of levels. 

By this approach, global tendencies of the experimental
low-lying spectra of $^{172-200}$Pt nuclei are reproduced quite well not
only for ground-state but also for side bands of mainly open-shell nuclei. 
It has been shown that shape/phase transition occurs quite smoothly from
prolate to oblate deformations as a function of $N$ in the 
considered nuclei $^{186-192}$Pt, where the $\gamma$ instability plays
an essential role. 
From the analysis in Fig.~\ref{fig:pes}, the change
of the mapped IBM PESs in $\gamma$ direction has been more vividly seen 
than in $\beta$ direction, similarly to the corresponding Gogny-HFB 
PESs of Ref.~\cite{RaynerPt}. 
This is consistent with the conclusions in our 
earlier work \cite{RaynerPt} and also with many others along the same
line. 
We have shown that the 
calculated spectra and the $B$(E2) ratios behave consistently with the
evolution of the topologies of the mapped PES's and with the systematics
of the derived IBM parameters as functions of the neutron number $N$. 
These derived parameters are qualitatively quite similar to the existing 
phenomenological IBM studies \cite{CastenCizewski,BijkerOs}. 
By studying the level schemes in detail in comparison with the available
experimental data, the present calculation agrees with the data fairly
nicely and reflects the algebraic aspects of the IBM, e.g., the
$\Delta\tau=\pm 1$ selection rule of the E2 decay patterns. 
We have also made predictions on some E2 transition patterns. 
These behaviors of the $B$(E2) may need to be 
examined experimentally particularly for lighter, 
$A\lesssim 190$ nuclei with which there is currently few available data. 

The evolution of ground-state shape as a function of 
both $N$ and $Z$ has been studied within neighboring isotopic chains
such as Os, W, Hf and Yb \cite{gradient-2}. 
More systematic analysis is in order for these nuclei, by more extensive
application of the present approach. 
It should be then of interest to study how the corresponding spectra and 
transition probabilities behave. 

On the other hand, the IBM Hamiltonian of Eq.~(\ref{eq:bh}) has 
rather simple form consisting of single-$d$-boson operator and the 
quadrupole-quadrupole interaction between proton and neutron bosons. 
The results provided by the present Hamiltonian were shown to be
already quite promising. 
However, more studies may be necessary in the future for further
refinement, e.g., in describing detailed structure of the quasi-$\gamma$
band. 
Work along this line is in progress. 

\section*{Acknowledgments \label{sec:acknowledge}}
\addcontentsline{toc}{chapter}{Acknowledgments}

We thank D.~Vretenar for valuable discussions. 
This work has been supported in part by Grant-in-Aid for Scientific
Research (A) 20244022 and by Grant-in-Aid for JSPS Fellows (No.~217368). 
Author K.N. is supported by JSPS Fellowship program. 
Part of this work has been carried out during his visit to the European 
Center for Theoretical Nuclear Physics and Related Areas (ECT*). 
He acknowledges ECT* and A.~Richter for their kind hospitality. 
The work of authors L.M.R and P.S has been supported by MICINN (Spain) under
research grants 
FIS2008--01301, FPA2009-08958, and FIS2009-07277, as well as by 
Consolider-Ingenio 2010 Programs CPAN CSD2007-00042 and MULTIDARK 
CSD2009-00064. Author R.R. acknowledges the support received within the
framework of the FIDIPRO program (Academy of Finland and University of
Jyv\"askyl\"a) and thanks Profs. 
J. \"Aysto and R.Julin as well as the  experimental 
teams of the University of Jyv\"askyl\"a (Finland) for 
warm hospitality and encouraging discussions.

\end{document}